\documentstyle[12pt]{article}
\begin{document}
\vfill
\centerline{\Large The Origin of Non-chaotic Behavior in Identically Driven
Systems.
}
\vspace{10pt}
\centerline {\bf P. M. Gade and Chaitali Basu}
\vspace{3pt}
\begin{center}
International Centre for Theoretical Physics\\
P.O. Box 586, Trieste 34100, ITALY\\
\vskip .5cm
\end{center}
\vspace{6pt}
\begin{abstract}
	Recently it has been found that different physical systems
driven by {\it identical} random noise behave exactly identical
after a long time. It is
also suggested that this is an outcome of finite precision in numerical
experiments. Here we show that the origin of the non-chaotic behavior
lies in the structural
instability of the attractor of these systems which changes to a stable
fixed point
for strong enough drive. We see this to be true in all the systems
studied in literature. Thus we affirm that
in chaotic systems, synchronization  can not occur only by addition of
noise unless the noise destroys the strange attractor and the system
is no longer chaotic.
\end{abstract}

Of late there has been considerable attention drawn to the problem
in which the behavior of an ensemble of initial conditions obeying
the same laws of
motion and driven by an identical sequence of random forces
start following the same random trajectory asymptotically.

This is observed in several  distinct cases. Fahy and Hamann \cite{Fah}
observed that if a particle obeying Newton's equations (without friction)
in a potential $V$ is stopped at regular time intervals $\tau$ and
all its velocity components are reset to random values, then for a
given sequence  of random values asymptotic trajectories become
identical for all initial conditions. This a stronger statement than
the observation that the statistical distribution of trajectories
become independent of initial conditions. This phenonomenon
occurs for the choice of a time interval $\tau$ lower than a threshold value
$\tau_c$. They remark that this is common for all
bounded systems.

Another observation of a similar phenomenon is due to Maritan and Banavar
\cite{Mar}. In this letter, they show that a pair of chaotic systems
driven by  identical noise of sufficient strength have identical
trajectories asymptotically. They studied the noisy logistic map and
 the Lorenz system. They have also given an analytic argument for the
above phenomenon in terms of Langevin dynamics \cite{Mar}. It was pointed
out by Pikovsky \cite{Pik} that the phenomenon of synchronization in logistic
maps does not occur in numerical experiments in quadrupole precision.
Thus he has suggested that the synchronization in systems with positive
Lyapunov exponent is a
numerical effect of the insufficient precision of calculations.
Maritan and Banavar \cite{Marp} replied that a pair  of chaotic
system subjected to the same noise has higher probability of becoming
identical compared to the case in absence of noise.

The observation by Pikovsky  raises some fundamental questions.
The question is whether all
the above reported observations are indeed physical phenomena or just an
artifact of finite precision of calculations in which two trajectories
that come close within the precision of numerical calculations
stick together. This question is important since there has been an
attempt to understand  systems like globally coupled maps
using maps driven by identical random noise \cite{Sin}.
These systems have  more experimental relevance \cite{Kan}.
In this system of globally coupled maps, it is observed that
though the maps form clusters at a particular
precision, the clustering disappears at higher precision \cite{Kan}.
This clustering is also observed in globally coupled Josephson Junction
arrays (JJA) \cite{Dan}. It is difficult to know whether this
clustering in JJA will change
with changing precision as the original calculations themselves are in double
precision and it is difficult to demand higher accuracy.
In short, it questions the reliability of numerical
computations in this and similar phenomena which are of higher
physical relevance.

One more question that arises in this context is how vital
is the presence of randomness for such synchronisation?

In this letter, we show that this synchronization phenomenon
is indeed physical in certain cases. It is an effect of
the fact that the strange attractor of the undriven
system is replaced by
a stable fixed point for strong enough perturbation. When the
attractor is a stable fixed point, different trajectories converge
exponentially towards it. If the parameter is now driven randomly
with the system flipping between chaotic and non-chaotic
regimes, and if the overall rate of contraction
of the trajectories is higher than that of the divergence in
the chaotic regime, the trajectories will converge for all precisions.
Here we point out that the randomness is of marginal significance and
even if the drive is periodic or constant in nature
the phenomenon of synchronisation occurs.

 We stress that the statement of 'synchronization of
chaotic systems subject to same random noise' is not correct
as the trajectories no longer remain chaotic. We illustrate this
statement by the example of the Lorenz system.

The numerical experiment in \cite{Fah} can be formulated in one-dimension
as
\[
x((n+1)\tau)=G(x(n \tau),v(n\tau))
\]
\begin{equation}
v(n\tau)=r_n
\label{F}
\end{equation}
where $r_n$s are delta correlated random numbers chosen from some
distribution and
$G(x,v)$ is a function which gives the final position of the particle
starting from $(x,v)$ and undergoing motion in the potential $V(x)$
for time $\tau$.

To check the essentiality of randomness in this example
we set all $r_n$s to a constant $\mu$, for the one dimensional
Duffing potential  $V(x)=x^4-x^2$ as studied in \cite{Fah}.
We observe that for smaller values of $ \mu $, there is
a high probability that the  system relaxes to a stable fixed
point for $\tau= 0.5$. In other words the
point attractor at the fixed point of $G$, is relatively
stable for smaller values of $\mu$, though there is no clear
critical point below which the fixed point is always stable.
We define a range
$I_s=[-\mu_c(\tau), \mu_c(\tau)]$,
such that the trajectories do not converge to a fixed point outside
this range.
In Fig. 1,
we give a plot of the coarse grained probability that
a fixed point is stable as a function of $\mu$ for various values of $\tau$.
As can be seen from the plot, it is more and more unlikely to have
a stable fixed point for higher values of $\mu$. This trend is more pronounced
for higher values of $\tau$. Now it is clear that if one chooses the
$r_n$s from the range of values for which there is high probability that the
fixed point is stable, the trajectories will converge. The fact that
randomness is of marginal significance is checked by putting
$r_n=4 \; \cos(\delta \; n),\; \delta =0.001$ for $\tau=.5$
where we still observe the convergence of trajectories ($\mu_c(.5)\simeq 3.7$).
This is  due to the fact that the fixed point
changes only continuously with $\mu$ and the trajectories keep on getting
attracted exponentially.
The range $I_s$ will also determine the average rate of convergence.
However, even if the the overall rate diverges, but the
the random number $r_n$ takes succesive values in the
regime  $I_s$ for a long enough time, the trajectories may
come close together within numerical precision and virtually
converge. It may not be possible to see this convergence in
numerical experiments with higher precision and thus
in these cases one may observe a precision
dependent convergence.

Here we point out the importance of the requirement of {\it boundedness}
in these systems. In bounded systems the existance of the fixed point
for $G(x,v)$ in Eq. \ref{F} with $r_n$s set to constant $\mu$ is
guaranteed \cite{7}. Now the question is why it
 is always stable for small values of $\tau$? It can be argued the
following way. The trajectories move towards the minimum of the
potential energy with some of the potential energy getting
converted into kinetic energy.
The successive reassignment of the velocity to a reasonably small value
after each time interval $\tau$ forces the system to dissipate some
part of the kinetic energy and hence also the total energy. This trend
continues till the particle comes near
the potential minimum. Here the motion of the particle gets
bounded within a part of the phase space near the potential
minimum. The fixed point $x_0(\mu)$ must be in this bounded region.
The stability of this point is given by ${dG(x_0(\mu),\mu) \over
 {dx}} =
1-{d^2V \over{dx^2}} {\tau^2 \over{2}} $ in the small time limit.
This quantity is less than unity if the second derivative of the
potential is positive which is certain if
the fixed point is near the potential
minimum. (Potential minimum is a stable fixed point for the case of
zero velocity for any choice of $\tau$ and since the position of the
fixed point $x_0(\mu)$
as well as ${dG(x_0(\mu),\mu) \over  {dx}}$ vary smoothly with
$\mu$, we expect the stability to be retained for small enough
velocities.) Thus we explain the stability of the fixed point for small
enough times and velocities. However for large times and velocities
the behavior that is true at the potential minimum may not
continue. One can easily see this for an
exactly solvable case of harmonic oscillator where $G$ is known
and it can be checked that there is no critical time $\tau_c$
except that for the choice of $\tau$s such that $\omega\tau
=n\pi$, the fixed point becomes marginally stable and convergence of
trajectories is not possible.
In this case of harmonic oscillator, ${dG(x_0,\mu_0)\over{dx}} \leq 1$ always.
This explains the factors determining the
critical time $\tau_c$ above which the trajectories no longer converge.

Now we consider the example considered in \cite{Mar}. The
Lorenz system \cite{Ab} is considered which is defined as
\begin{equation}
\begin{array}{l}
{{dx} \over {dt}}=P(y-x)=f(x,y,z)\\
{{dy} \over {dt}}=-xz+mx-y=g(x,y,z)\\
{{dz} \over {dt}}=xy-bz=h(x,y,z).
\end{array}
\end{equation}
with P=10, b=8/3 and m=28, which is integrated with
time step of 0.001 unit. The $y$ equation is evolved as
\begin{equation}
y(t+\Delta t)= y(t)+ \int_t^{t+\Delta t} g(x,y,z) dt + r_t W \sqrt{\Delta t}.
\label{chg}
\end{equation}
where $r_t$ is an uniform random number in range $[0,1]$. For higher
accuracy we have chosen to use Runge-Kutta method with adoptive stepsize
control instead of Euler scheme used in ref. \cite{Mar}. For this
system, Ref. \cite{Mar} claims
that above a critical value of $W=W_c=2/3$ the synchronisation
occurs.

As in the case of \cite{Fah}, here also we check the necessity of randomness
for the synchronisation by replacing
the random number $r_t$ by unity and try to check the behavior for
different $W$s. We see that for $W> W_+\simeq 0.16$ or for
$W<W_- \simeq -0.16$, the
fixed point is stable. This can be checked by following it using Newton-Raphson
method and calculating eigenvalues and also by perturbing a little about
the fixed point and checking whether the system relaxes back to it.
Thus one has a stable
chaotic behavior only in the range $I_c=[W_-,W_+]$. Now in case when
the system is driven out of this regime $I_c$ by random forces, the
effective lyapunov exponent of the system becomes negative. As in
earlier case, it will depend on the time that the trajectories spend in
the region $I_c$ and outside it. In this case,
the trajectories can clump together if the lyapunov exponent is
negative. We point out that the
negative lyapunov exponent is necessary for this clumping.

The transient time required for clumping grows as we decrease the
value of $W$. However, we observe that within our computation time
and precision the critical value of $W$ is much less than the
one quoted in the text as $W_c=2/3$ which could be the result of
more precise integration scheme and longer
computation times.
Although, this quantitative
correction is of marginal significance,
we would like to point out that the statement in
ref. \cite{Mar} that,
'However, the strange
attractor is {\it not} replaced by anything simple such as a fixed point
or limit cycle.' is misleading since the strange attractor is indeed replaced
by a simpler structure at least in the parameter range outside $I_c$.
We also point out that
the largest lyapunov exponent
from the linearised equations of the variables of the system,
computed along the noisy trajectory,  is
negative in the parameter regime of convergence.
Thus the trajectories are no longer chaotic in the regime of convergence.
To check that it is indeed so, we do computations for large enough times as
the system is nonstationary and short-time computation can give a spurious
indication that the lyapunov exponent has converged to a positive
value.

In Fig. 2 we plot the largest lyapunov exponent as a function of
constant $W$ calculated at time $t= 6.4\; 10^4$.
We observe a sudden drop around $W=.19$ at which  the lyapunov exponent
becomes negative. This value could be slightly lower for larger computation
times. This explains the clumping in case when the random number in
Eq. \ref{chg} has higher mean, i.e. the value of $W$ is higher.

It is mentioned in \cite{Mar} that the phenomenon is not observed
if the random numbers $r_t$s are symmetric around 0. This assertion is  quite
correct since the $y$ component of stable fixed point changes its sign with
the sign of $W$ and there is a discontinuity in its stability
in $I_c$ regime. We note here that the stability of the fixed
point dictates the behavior of trajectories only in its close vicinity.
In this local neighborhood the trajectories get attracted to the fixed point.
However, this picture will not be true for the points that are not in its
neighborhood.  Thus this local convergence will not be valid for
the points which are away from the fixed point. The fixed point
is changing discontinuously and considerably in this case. Thus
due to the fact that the perturbation is being continuously
applied, trajectories may never come close enough to either fixed points
simultaneously.
This is the reason why one will not observe the collapse of trajectories
whereas it is possible in the case when the stable fixed point changes
only continuously.

We measure the lyapunov exponent for the synchronised case and see
that it is indeed negative in the regime in which synchronisation is
observed.  Thus the trajectories are no longer chaotic.

Here we also point out that similar situation can not arise in systems
in which such structural instability is not expected. However,
the case of precision dependent synchronization in logistic maps
has a different origin.
In \cite{Mar}, they have given an example of noisy logistic maps
to illustrate their point of view. However, when we repeated the
similar experiment for the case of noisy tent maps, we did
not observe this effect. This is an artifact of the fact
that tent map does not have locally negative lyapunov exponent at any
point\cite{own1}.

This synchronisation phenomenon is also observed in Ref \cite{lu}
in which they showed
that in a random dynamical system  variation of a parameter causes
a transition from a situation where an initial cloud of particles
eventually permanently clumps at a point to a situation where the
particles are eventually distributed on a fractal.

We would point out that the results in \cite{lu} do not clash with
our observations. Here the collapse of trajectories occur when
the lyapunov exponent is negative and this is expected.
They study the following two dimensional model
\[
x_{n+1}=[x_n+(1-e^{-\alpha}) \alpha^{-1} y_n]mod(2 \pi)
\]
\begin{equation}
y_{n+1}=e^{-\alpha}y_n+k \sin (x_{n+1}+\theta_n)
\end{equation}
where $\theta_n$s are independent random variables with uniform
probability density in $0\leq\theta\leq 2\pi$.
In this model, for the parameter range
in question, the fixed point is always stable in the absence of noise.
If one replaces noise by a constant still the same behavior is
observed except that the position of the fixed point changes.
However, for lower values of the parameter $\alpha$ dictating the
dissipation rate,
the continuous random forcing kicks the system out
of the basin of attraction of the fixed point and the
trajectories never clump together. Thus though the trajectories go to the
fixed point always in the absence of noise, for
noise of strength comparable to the dissipation, the trajectories
do not clump. Here dissipation forces the system
towards the fixed point and randomness drives it away
from it. These act as competing  parameters.
Thus it is possible that the trajectories that would
otherwise converge do not converge owing to random noise. We point
out that in the limit of strong dissipation the parameter range in
which the clumping occurs is the same as the parameter range in
which the fixed point is stable \cite{lu2}.

However, in this letter
we discuss the reverse phenomenon, namely of chaotic trajectories
converging
by random perturbations. We show that it  is possible only if
these perturbations
change the basic nature of the system and one has a stable nonchaotic
attractor for higher perturbations.

In conclusion, we have tried to trace the origin of the phenomenon
of clumping observed in different systems which are driven by identical
random forces. We have shown that it is due to the structural
instability in the system which results from large perturbations.
We have also tried to explain the conditions under which the precision
dependent clumping could occur. Here we note that the trajectories
converge and clump since these perturbations force it to a state
in which system has a negative lyapunov exponent. Thus the
system is no longer chaotic. The important
factors are dissipation and forcing whereas
randomness does not play a significant role. We have also discussed
the reverse phenomenon, namely the one in which the trajectories that would
otherwise go to the stable fixed point display positive lyapunov exponent.
Thus we do not see any clumping provided the
noise is high and dissipation is comparatively low.

We acknowledge discussions with Prof. Pikovsky, Prof. Maritan and Prof.
Cerdeira and a helpful correspondence with Prof. Banavar. We also thank
Prof. Pikovsky for sharing \cite{Pik} prior to publication.

\newpage
{\large Figure Captions}

\begin{itemize}

\item[Fig. 1]
The coarse grained probability that the trajectories converge to a
fixed point as function of velocity $\mu$ (i.e. setting $r_n=\mu$
for Eq. \ref{F}) for different values of $\tau$. The curves from
above to below represent $\tau=.005,.05,.1,.2,.3,.4$ and $.5$
respectively.
\item[Fig. 2]
The largest lyapunov exponent $\lambda$ as a function of $W$.
\end{itemize}

\begin{thebibliography}{99}
\bibitem{7} Since at $\infty$, $x((n+1)\tau) <x(n\tau)$ and at $-\infty$
$x(n\tau) <x((n+1)\tau)$, the continuous function $G$ must have
$x(n\tau)=x((n+1)\tau)$ for at least one point.
\end{thebibliography}
\end{document}